\documentclass[modern]{aastex63}

\usepackage{graphicx}
\usepackage{color}
\usepackage{amsmath}  
\usepackage{verbatim}     
\usepackage{hyperref}
\usepackage[noabbrev,capitalise]{cleveref}  
\usepackage{natbib}
\usepackage[mathscr]{euscript} 
\usepackage{xspace}
\usepackage[disable]{todonotes}
\usepackage{csquotes}
\usepackage{geometry}
\makeatletter
\usepackage{etoolbox}
\patchcmd\H@refstepcounter{\protected@edef}{\protected@xdef}{}{}
\makeatother

\newcommand{\sn}{SN Ia\xspace}
\newcommand{\sne}{SNe Ia\xspace}

\newcommand{\h}{\ensuremath{\text{H}_0}\xspace}

\newcommand{\un}[1]{~\text{#1}\xspace}  

\begin{document}

\title{\vspace{-2.5em}Synergies between Vera C. Rubin Observatory, \textit{Nancy Grace Roman Space Telescope}, and \textit{Euclid} Mission: Constraining Dark Energy with Type Ia Supernovae}
\shorttitle{Supernova Cosmology using Rubin, Roman, and Euclid}

\author[0000-0002-1873-8973]{B. M. Rose}
\affiliation{Department of Physics, Duke University, 120 Science Drive, Durham, NC, 27708, USA; \url{benjamin.rose@duke.edu}}
\author{G. Aldering}
\affiliation{Physics Division, E.O. Lawrence Berkeley National Laboratory, 1 Cyclotron Rd., Berkeley, CA, 94720, USA}
\author[0000-0002-5995-9692]{M.~Dai}
\affiliation{Department of Physics and Astronomy, Johns Hopkins University, Baltimore, MD 21218, USA}
\author[0000-0003-2823-360X]{S. Deustua}
\affiliation{Space Telescope Science Institute, 3700 San Martin Drive Baltimore, MD 21218}
\author[0000-0002-2445-5275]{R.~J. Foley}
\affiliation{Department of Astronomy and Astrophysics, University of California, Santa Cruz, CA 95064, USA}
\author[0000-0001-6728-1423]{E.~Gangler}
\affiliation{Laboratoire de Physique de Clermont, IN2P3/CNRS, Université Clermont Auvergne, F-63000 Clermont-Ferrand, France}
\author[0000-0001-5673-0959]{Ph.~Gris}
\affiliation{Laboratoire de Physique de Clermont, IN2P3/CNRS, Université Clermont Auvergne, F-63000 Clermont-Ferrand, France}
\author[0000-0002-2960-978X]{I.~M. Hook}
\affiliation{Department of Physics, Lancaster University, Lancaster, LA1 4YB, U.K.}
\author{R. Kessler}
\affiliation{Kavli Institute for Cosmological Physics, University of Chicago, Chicago, IL 60637, USA}
\affiliation{Department of Astronomy and Astrophysics, University of Chicago, Chicago, IL 60637, USA}
\author[0000-0001-6022-0484]{G. Narayan}
\affiliation{University of Illinois at Urbana-Champaign, 1003 W. Green St., IL 61801, USA}
\affiliation{Centre for Astrophysical Surveys, National Centre for Supercomputing Applications, Urbana, IL 61801, USA}
\author[0000-0002-3389-0586]{P. Nugent}
\affiliation{Physics Division, E.O. Lawrence Berkeley National Laboratory, 1 Cyclotron Rd., Berkeley, CA, 94720, USA}
\affiliation{Department of Astronomy, University of California, Berkeley, CA 94720-3411, USA}
\author{S. Perlmutter}
\affiliation{Physics Division, E.O. Lawrence Berkeley National Laboratory, 1 Cyclotron Rd., Berkeley, CA, 94720, USA}
\affiliation{Department of Physics, University of California Berkeley, 366 LeConte Hall MC 7300, Berkeley, CA, 94720, USA}
\author[0000-0002-8207-3304]{K.~A.~Ponder}
\affiliation{SLAC National Accelerator Laboratory, 2575 Sand Hill Rd, Menlo Park, CA 94025, USA}
\author[0000-0001-8861-3052]{B.~Racine}
\affiliation{Aix Marseille Univ, CNRS/IN2P3, CPPM, Marseille, France}
\author[0000-0001-5402-4647]{D. Rubin}
\affiliation{Physics and Astronomy Department, University of Hawaii, Honolulu, HI 96822, USA}
\affiliation{Physics Division, E.O. Lawrence Berkeley National Laboratory, 1 Cyclotron Rd., Berkeley, CA, 94720, USA}
\author[0000-0002-8687-0669]{B. O. S\'anchez}
\affiliation{Department of Physics, Duke University, 120 Science Drive, Durham, NC, 27708, USA; \url{benjamin.rose@duke.edu}}
\author[0000-0002-4934-5849]{D. M. Scolnic}
\affiliation{Department of Physics, Duke University, 120 Science Drive, Durham, NC, 27708, USA; \url{benjamin.rose@duke.edu}}
\author[0000-0001-7113-1233]{W.~M.~Wood-Vasey}
\affiliation{
Pittsburgh Particle Physics, Astrophysics, and Cosmology Center (PITT PACC).
Physics and Astronomy Department, University of Pittsburgh,
Pittsburgh, PA 15260, USA
}

\author[0000-0001-5201-8374]{D. Brout}
\altaffiliation{NASA Einstein Fellow}
\affiliation{Center for Astrophysics, Harvard \& Smithsonian, 60 Garden Street, Cambridge, MA 02138, USA}
\author[0000-0001-7101-9831]{A. Cikota}
\affiliation{European Southern Observatory, Alonso de Cordova 3107, Vitacura, Casilla 19001, Santiago de Chile, Chile}
\affiliation{Physics Division, E.O. Lawrence Berkeley National Laboratory, 1 Cyclotron Road, Berkeley, CA, 94720, USA}
\author{D.~Fouchez}
\affiliation{Aix Marseille Univ, CNRS/IN2P3, CPPM, Marseille, France}
\author[0000-0003-4069-2817]{P. M. Garnavich}
\affiliation{University of Notre Dame, Center for Astrophysics, Notre Dame, IN 46556, USA}
\author[0000-0002-0476-4206]{R. Hounsell}
\affiliation{University of Maryland, Baltimore County, Baltimore, MD 21250, USA}
\affiliation{NASA Goddard Space Flight Center, Greenbelt, MD 20771, USA}
\author[0000-0003-2764-7093]{M.~Sako}
\affiliation{Department of Physics and Astronomy, University of Pennsylvania, 209 South 33rd Street, Philadelphia, PA 19104}
\author{C.~Tao}
\affiliation{Aix Marseille Univ, CNRS/IN2P3, CPPM, Marseille, France}
\author[0000-0001-8738-6011]{S.~W.~Jha}
\affiliation{Department of Physics and Astronomy, Rutgers, the State University of New Jersey, 136 Frelinghuysen Road, Piscataway, NJ 08854, USA}
\author[0000-0002-6230-0151]{D.~O.~Jones}
\altaffiliation{NASA Einstein Fellow}
\affiliation{Department of Astronomy and Astrophysics, University of California, Santa Cruz, CA 95064, USA}
\author[0000-0002-7756-4440]{L.~Strolger}
\affiliation{Space Telescope Science Institute, 3700 San Martin Drive Baltimore, MD 21218}
\author[0000-0003-1899-9791]{H. Qu}
\affiliation{Department of Physics and Astronomy, University of Pennsylvania, Philadelphia, PA 19104, USA}

\shortauthors{B. M. Rose, et al.}

\date{March 8, 2021}
\reportnum{\hspace{-10em}DOE/NASA Request for Information: Focus Area 3}

\begin{abstract}
We review the needs of the supernova community for improvements in survey coordination and data sharing that would significantly boost the constraints on dark energy using samples of Type Ia supernovae from the Vera C. Rubin Observatories, the \textit{Nancy Grace Roman Space Telescope}, and the \textit{Euclid} Mission.  We discuss improvements to both statistical and systematic precision that the combination of observations from these experiments will enable. For example, coordination will result in improved photometric calibration, redshift measurements, as well as supernova distances. We also discuss what teams and plans should be put in place now to start preparing for these combined data sets. Specifically we request coordinated efforts in field-selection and survey operations, photometric calibration, spectroscopic follow-up, pixel-level processing, and computing. These efforts will benefit not only experiments with Type Ia supernovae, but all time-domain studies, and cosmology with multi-messenger astrophysics.
\end{abstract}

\NewPageAfterKeywords

\section{Introduction}

Type Ia Supernovae (\sne) are a key probe for measuring dark energy, and currently hold a critical place in two of the three large surveys: the Legacy Survey of Space and Time (LSST) with the Vera C. Rubin Observatories and \textit{Nancy Grace Roman Space Telescope}.  Both LSST and the \textit{Roman} \sn surveys will be unprecedented in their statistical sample size.  The LSST optical sample at redshift $0<z<1$ will be a factor of $300\times$ larger than the cumulative sample of \sne today and the \textit{Roman} NIR sample at $0.3<z<3.0$ will discover and measure $50\times$ more \sne at $z>1$ than in our sample today and increase the sample of NIR light-curves by a similar magnitude.  While \textit{Euclid} does not have a planned \sn survey in its baseline plan (see here for a proposed strategy: \citealt{Astier2014}), it will revisit deep fields with a regular cadence, which can produce \sn detections and be combined with the LSST sample.  

The synergy between combined data sets is enormous --- especially if the surveys coordinate observations.  Combining data sets allows increases in redshift range and wavelength range, boosting the statistical precision and the systematic control. Furthermore, as discussed in other responses to this call, the \sn programs will immensely benefit from cross-survey data products to improve calibration and photometric redshifts. Particularly due to the time-dependent nature of \sn surveys, planning must be done now for survey coordination and combined analyses. In this response to the DOE/NASA Request for Information, we separate aspects that need decisions now versus those where planning should begin soon to be considereded for future synergies.

\vspace{.1in}

For the \sn programs, it is particularly important that the {\it multiple} science agencies who have issued this RFI, continue to work together because the cosmological analyses will be built on \sn data that must be obtained using large time allocations from facilities that are run by these (and other) agencies.   The quality of the results will also strongly depend on the development of techniques and knowledge that individually may go beyond each different agency's traditional focus areas. To maximize the science impact of each agency's investment, scientists must have broad support, including for ways that allow them to work together across agencies' traditional boundaries. 
A scientist should only need to  apply to a single agency for funding to fully participate in all aspects of the problem.

\section{Science Enhancements}

\begin{displayquote}
\textbf{a. What are the key dark energy science areas that will be enhanced by these activities? What level of scientific enhancement is expected by carrying them out after the datasets are public? 
}
\end{displayquote}

The LSST and \textit{Roman Space Telescope} surveys will produce \sn samples with significantly different, but overlapping redshift ranges.  LSST will discover an enormous ($>${}$300,000$) number of SNe around $z\approx0.3$--$0.4$ in its wide survey, and a smaller ($>${}10,000) but more cosmologically-constraining number in deep fields out to $z\approx0.9$--$1.0$. \textit{Roman} will discover a smaller number ($>${}$10,000$) but reaching to much higher redshifts ($z > 2.0$).  Combining \sn samples from LSST and \textit{Roman} improves the dark energy figure of merit \citep{Albrecht2006} by up to a factor of two, over independent analyses. In addition, there will be hundreds of strongly gravitationally lensed \sne from these surveys in which joint analysis will greatly improve their cosmological utility in a similar manner as it will for the normal \sn population \citep{2019ApJS..243....6G}.

With such high statistical precision, analyses will leverage systematic uncertainty `self calibration' \citep{Kim2006, Rubin2015, Brout2020b}, but will nonetheless be limited by systematic uncertainties.  Combining data from Rubin, \textit{Roman}, and \textit{Euclid} --- as well as complementary spectroscopic follow-up --- will greatly improve control of certain systematic uncertainties.  There are multiple reasons for this improvement due to both better instrumental  calibration and improved knowledge of \sn physics.  A brief list of the systematic uncertainties that can be improved are:
\begin{itemize}

\item \textbf{Photometric calibration} A key source of systematic uncertainty for measuring the dark energy equation-of-state is the lack of a common, accurate, chromatic scale referenced to the physics-based International System (SI) The uncertainty in the flux standards, and hence, uncertainty in the photometric zero points are a large contributor to the systematic error budget \citep{Hounsell2018, cosmicvisions}.  Work to establish such calibration is a focus of several groups --- for example,  STarDICe, SCALA, NISTStars, Collimated Beam Projector, ORCASat, ORCAS --- that are not-accidentally dominated by \sn researchers \citep{2017A&A...607A.113L, Aldering2021, 2018SPIE10704E..20C}. Measurements of dark energy, in particular \sn cosmology, require an external calibration error budget, so we must rely on more than one of these paths to be successful.  Between them, \textit{Roman}, \textit{Euclid}, and LSST  span the optical and near-infrared wavelengths, presenting a special challenge.  They require not only joint observations of SI-referenced optical-NIR spectrophotometric standard stars, but also must ensure photometric consistency between these experiments through coordinated observations and measurement algorithms for agreed-upon calibration fields.
One key activity for achieving the desired cross-mission accuracy  is for each mission/experiment to commit the necessary  resources towards  understanding and tracking instrument performance over time, in addition to  wavelength.

\item \textbf{Spectroscopic follow-up} Combining data from Rubin, \textit{Roman}, \textit{Euclid}, and complementary spectroscopic follow-up will greatly improve control of certain systematics. Systematic biases in SN~Ia distances can be controlled (or in some cases eliminated) with spectroscopic observations of the supernovae themselves.  These observations can remove residual non-Ia contamination and mis-identified redshifts, minimize host-galaxy SN-luminosity dependence, test for population evolutionary drift, and significantly improve the statistical uncertainty of the standardization \citep{Jones2018,Boone2021}.

\item \textbf{Photometric redshift measurements} are improved with a broader wavelength coverage \citep{Capak2019,Rhodes2019}. Photometric redshifts will be an important piece of the overall strategy to obtain \sn redshifts. The training of photometric redshift algorithms will also be improved by synergies between observatories, see the response to this RFI ``Photometric Redshifts for the Next Generation of Weak Lensing Surveys'' by Daniel Masters, et al. for details.

\item \textbf{Host-galaxy redshifts} In order to measure dark energy with \sne, redshifts are needed. These can come from the SN or its host galaxy. LSST-DESC has baselined obtaining redshifts from host-galaxies due to the multiplex advantage of this approach, and discussions are ongoing to partner with 4MOST for this purpose. But 4MOST will not reach sufficient depth in the LSST Deep-Drilling Fields (DDF), where most of the DESC \sn cosmology will be performed. \textit{Roman} will estimate a fraction of its redshifts from the spectra obtained with its prism and grism surveys. \textit{Roman} also has access to 100 nights of time on Subaru, where the PFS instrument could be used to obtain these redshifts for \textit{Roman} SN fields north of declination $-30^\circ$. Simulations by the \textit{Roman} SN SITs show that more than 100 nights on Subaru with PFS could be productively consumed by the \textit{Roman} SN program alone, because there will be a sizable population of intrinsically faint host galaxies of high-redshift \textit{Roman} \sne. Therefore, there is certain to be an unmet need for additional SN host redshifts. A coordinated effort would result in a more efficient follow-up campaign.

\item \textbf{\sn standardization} Systematic biases in \sn distances are better understood when combining light-curve data across broad wavelengths, such as LSST+ NIR from Rubin and \textit{Euclid}~\citep{Ponder2020,Uddin2020}. These observations would need to be coordinated so all surveys could observe the same transients. While rest-frame optical (and rest-frame NIR) bands are highly correlated among themselves, optical and NIR bands are relatively uncorrelated, indicating unique information when observing across this wide wavelength range \citep{Mandel2011}. Additionally, a broader wavelength range provides a larger lever arm to constrain dust reddening and perhaps disentangle the effects of intervening dust from intrinsic color \citep{Brout2020, Mandel2020}.  In addition, \sne have a lot of diversity in the rest-frame UV, which can help to perform better standardization if it can be exploited \citep{Ellis2008,Cooke2011,Foley2012a,Leget2020, Boone2021}. The complimentary observed wavelengths of \textit{Roman} and Rubin allow for improved \sne standardization at a broader redshift, resulting in better constraints on dark energy.  Furthermore, many studies have shown that properties of SNe correlate with properties of the host-galaxy properties, and these correlations can be used to better standardize SNe~Ia.  Combined photometry from the different missions will allow for greater wavelength range to determine host-galaxy properties and significantly improve these studies.

\item \textbf{Selection functions} When combining surveys that target different redshifts, one can better characterize selection effects and faint source photometry by comparing photometry of different surveys with different depths \citep{Rubin2015, Kessler2017}.

\end{itemize}

\begin{displayquote}
\textbf{b. What is the scope of work required, as well as the opportunities and costs?}
\end{displayquote}

There are a number of work items that fall under the theme of ``survey coordination.''
Due to the timeline of LSST, much of this needs to be done now.  A TAG survey coordination task-force is already in place, though their charge has been mainly limited to sky-area selection (deep and wide fields). While there are currently groups in place to discuss survey strategy coordination, a key obstacle is different timelines between when decisions need to be made by each survey. What needs to be coordinated now:
\begin{itemize}
    \item A coordinated selection of overlapping deep field location, time, season length, and filters.
    \item Establishment of an all-sky calibration network: to make sure all surveys can observe standard stars from same network.  This includes ensuring that SI-traceable, calibrated standard stars are available at multiple locations around the sky, as well as establishing a set of ``touchstone" fields suitable for survey facilities.  Stable stars in these fields should be identified, and if possible, subjected to the same vetting process as the traditional, individual flux standards, as in the CALSPEC \citep{Bohlin2014} data base. There are have been a few  independent efforts, the most recent and extensive being faint, northern  white dwarf standards placed on the CALSPEC system \citep{Narayan2019, Calamida2019}. 
     
\end{itemize}
What needs to be planned soon:
\begin{itemize}
    \item Forced-position photometry on subtracted images from surveys with faint detection.
    \item Prioritized spectroscopic follow-up for supernovae and host-galaxies.
    \item Artificial {\it source} injection on real images for measuring detection efficiency and validating photometric pipeline. 
\end{itemize}

\textbf{Forced-position photometry on subtracted images.}
In terms of forced-position photometry, while all surveys have plans for forced photometry at specific locations, there is no plan in place for coordination across surveys.   There should be a system to share lists of SN positions and peak dates (for building templates).
Furthermore, it is unclear if \textit{Euclid} plans to perform image subtraction, and a separate agency/survey may be necessary to provide the software necessary to perform this task. Archive access, between science analysis platforms, is needed to allow reprocessing for consistency and homogeneity will be critical, particularly for transient detection and classifications. This would enable, for example, using SN locations determined from one survey to perform forced-position photometry on data from a second survey.

Consistent observation and modeling of \sne in their host galaxies will require forced-position photometry or scene modeling of pixel data based on recorded positions of detected supernovae.
Specifically, SNe detected by \textit{Roman} should be photometered in the Rubin data, even in cases where the SN is below the detection threshold.

\textbf{Follow-up spectroscopy} Due to the time-sensitive nature of transient science, combining information across surveys must be done quickly.  This is particularly important if one wants to include interesting transients or observe \sne early in their evolution.  For spectroscopic programs, there are multiple options including Subaru/PFS and The Dark Energy Spectroscopic Instrument (DESI). Priority should be given to targets in fields from multiple telescopes. DOE's DESI instrument could help fulfill this need for any \textit{Roman} fields or LSST DDF fields it can reach. For any blocks of the \textit{Roman} wider-field tier that are contiguous over more than 7~sq.~deg, DESI will be as efficient as PFS (in terms of host redshifts per hour). Trial observations with DESI -- requiring total exposure on a field in the neighborhood of 25~hr -- could test the power of DESI for this application. The result could influence the selection of the sky regions for the \textit{Roman} SN program. Field selection will take place over the next few years.

To facilitate the coordination of spectroscopic follow-up, multiple `Broker' groups are studying optimal prioritization.  This includes a joint effort between DESC and the Cosmostatistics Initiative (COIN) to develop a Recommendation System for Spectroscopic Follow-up~\citep[RESSPECT; ][]{Kennamer2020}, which will optimize follow-up taking into account multiple metrics including photometric classification uncertainty and availability of telescope time for spectroscopic follow-up. With collaboration between the three experiments, we can construct different metrics to optimize combined science goals.

\textbf{Artificial source injection} Artificial source injection on real images is used to measure single-visit detection efficiency vs.\ S/N in each bandpass \citep{Pain1996, Kessler2019}, measure efficiency vs.\ S/N for machine-learning used to reject subtraction artifacts \citep{Bailey2007, Godlstein2015}, and to characterize anomalous noise on bright galaxies (Sec.~6.4 of \citealt{Kessler2019}; Sec.~4.4 of \citealt{Brout2019smp}). The resulting efficiency and anomalous-noise maps are used to measure biases in simulated data to correct for expected biases in real data. Artificial source magnitudes should span from near saturation to $\sim 2 \un{mag}$ beyond the detection limit. Artificial sources need not be associated with realistic light curves, but should be placed near galaxies following their light profile. 
Ultimately, artificial light-curve injection is useful to check the entire cosmology analysis as shown in Sec.~6 of \citet{Brout2019}. From experience with this data set, artificial SN light curves provide an
adequate flux range to measure a single-visit detection efficiency.

Rubin and \textit{Roman} will have the bulk of their SNe at different redshifts.
Analyses that are sensitive to potential SN~Ia population drift and detection efficiency will be greatly improved and less total work if done consistently for both Rubin and \textit{Roman}. This can be done both on the real data once available, but should also be done on consistent, simulated images.
In particular, the part of the pixel processing that matters for SN sensitivity is the distribution of SN light on top of different host galaxy types and local surface brightness.
An effort to do this correctly and consistently across \textit{Euclid}, Rubin, and \textit{Roman} will (1) be much simpler to understand and compare across analyses; and (2) be less total effort to do jointly than for each team to re-do and then have to re-interpret across different teams.

\begin{displayquote}
\textbf{c. What are key obstacles, impediments, or bottlenecks to advancing development of these plans?}
\end{displayquote}

The key operational obstacles are communication and timeliness. Since \sne reach peak only a few weeks after discovery, communication must be done both proactively and in real time. Speed is even more important for ancillary science with rare transients where their evolution is unknown.  Furthermore, it is still unclear which data can be shared and with what latency.  This will complicate efforts to optimally yield the best cosmological constraints from joint observations and joint analyses. The key structural challenge is giving teams authority to work on topics across projects.
Many endeavors will work best when a team working on a topic under one project is encouraged to make the connection to complementary projects.
However, this will requires agencies to be willing and able to coordinate.

There are also some potential challenges with the proprietary nature of the Rubin Observatory LSST data.
The \textit{Roman} data are fully public, while the LSST data have a 2-year proprietary period.
The \textit{Roman} science collaborations will be international, and unlikely to precisely match up with international data rights in LSST.
Thus if one wanted to do a joint \textit{Roman}+LSST analysis of the first year of \textit{Roman} data, the LSST data on those same supernovae would not be available to the full \textit{Roman} community until 2 years later.
However, the LSST data rights community {\it could} do a full analysis of the LSST + \textit{Roman} data.

\begin{displayquote}
\textbf{d. Are there other science topics besides dark energy that drive the requirements for joint data processing or analysis?}
\end{displayquote}

Simultaneous observations with Rubin and \textit{Roman} will result in a large sample of transients beyond \sne.  Combined optical/NIR observations will constrain dust properties of extinguished supernovae and reveal other exotic transients whose emission peaks at $\sim$1~$\mu$m such as luminous red novae and kilonovae.  Core-collapse supernova rates, which constrain the cosmic star-formation rate, can be more accurately determined without requiring a dust-obscured correction.

The hundreds of strongly gravitationally lensed \sne observed from these surveys, though not a part of typical cosmological analyses, would benefit from joint analysis and result in a significant improvement to their cosmological utility \citep{2019ApJS..243....6G}. Furthermore, multi-wavelength follow-up of gravitational wave counterparts is necessary to constrain the composition of the ejected material.

\section{Collaboration and Partnerships}

\begin{displayquote}
\textbf{k. What cooperation or partnerships between DOE and NASA could further the scientific and technology advances?}
\end{displayquote}

\begin{itemize}
\item While a TAG Survey Coordination group is already formed (led by Dan Scolnic and Daniel Stern), communication between this group and the collaborations should be formalized and the charge expanded to improve survey operations to further coordination between survey planners, specifically with the authority and ability to tweak the strategies before and during the multiple surveys. We advocate that all survey strategies be reviewed and adjusted periodically during operations, in order to further optimize synergies based on what is learned from initial data sets, and based on what new observing facilities become available.

\item The creation of a joint survey calibration task force to establish communication across the missions, identify common algorithms, and ensure coordination so that uncertainties in photometric calibration are not the limiting factor in achieving Stage IV dark energy figure of merit \citep{Albrecht2006}. 

\item Spectroscopic follow-up task force to ensure that access to sufficient follow-up spectroscopy resources is obtained, through different TACs, MoUs with external groups in exchange for data rights, and through the construction of new facilities and instruments. There will be multiple types of spectroscopic follow up to achieve the different goals of SN identification, redshift determination, and statistical and systematic error constraints, and to follow SN targets that are visible in different hemispheres at different times.  Moreover, systems (and pipelines) will need to be planned, organized, and developed to ensure rapid and efficient use of all available spectroscopic resources.
As an example of a prioritization issue that this planning should address: if a SN is observed by multiple surveys, it should be preferentially allocated appropriate follow-up resources, while avoiding duplication of observations to maximize science return.

\item Expansion of in-place joint pixel-processing task force to generate a plan for all surveys to grant access of pixel-level data associated with transients, in less than one day turn round, and with oversight of the development of the necessary infrastructure to realize this goal. Many images will be crowded and confusion-limited, making catalog-only data of limited utility relative to the previous generation of ground-based surveys. A joint pixel-processing facility will be necessary to provide postage stamps generated from the images taken from the different surveys, together with a probabilistic source separation, and probability density functions of photometric redshift. The same facility will be useful for understanding the host environments of transients of all varieties, and in particular how galaxy evolution impacts transient explosions. Deep multi-survey images of the host will also be useful for early classification, and to detect rare strongly-lensed supernovae which will need prioritized followup. 

Joint pixel-processing of data from different surveys will also be necessary for multi-messenger astrophysics (e.g. detecting the optical counterparts of events seen in future surveys such as CMB-S4). Services and standards to deliver alerts for detections in multi-messenger missions such as \texttt{HopSkotch} are being developed by the SCiMMA group\footnote{\url{https://scimma.org/}}, and coordination between these groups would be a major responsibility of such a joint processing task force. A joint pixel processing facility is also crucial to several other probes of dark energy, detailed in a response to this RFI from LSST DESC in Annis et al., ``Response from the Rubin Observatory LSST Dark Energy Science Collaboration: Extending LSST Studies of Dark Energy and Dark Matter with \textit{Roman} and \textit{Euclid}''.

\item A real-time forced-position photometry task force to recover measurements for SN detected in other surveys. As missions detect SNe, it can be very valuable to recover sub-threshold measurements from other surveys. It is possible to study the earliest phases of a potentially strongly-lensed SNe detected in LSST several days after explosion if it is possible to extract forced photometry at the location of other lensed images from \textit{Roman} observations. A forced photometry service should also be capable of working together with a pixel processing facility to allow reprocessing of data, as well as fake source injection to model the joint survey selection function. This will be essential to understand the demographics of novel and rare classes of transients such as pair-instability supernovae, as well as to propagate measurements from each of these missions into joint cosmological constraints on the properties of dark energy.

\item Joint computational task force to manage shared computing of these multi-mission datasets, including common simulations (e.g. simulating the multi-mission sky to include the effect of a novel dark energy model, to ensure its effects are correctly propagated through all three surveys), and tools for data access and processing (e.g. a shared JupyterHub environment that scientists working on any of these missions can use). Computing is the central nexus between observers, theorists, and instrumentalists, and now encompasses astrostatistics and data science, in addition to computational physics. A research platform facility that colocates (or provides transparent access to) data holdings and provides computing will also be essential for multi-messenger astrophysics, which has a similiar need to combine observations from several different observatories (e.g. LIGO-VIRGO-Kagra, IceCube and ngVLA) together with \textit{Euclid}, \textit{Roman} and Rubin.
 
\end{itemize}

\begin{displayquote}
\textbf{l. What mix of institutions or collaboration models could best carry out the envisioned research and/or development?}
\end{displayquote}

We advocate the use of cross-survey (and agency) task forces to plan and coordinate the above synergistic activities. These task forces should have both sufficient scientific expertise and authority to be able to address specific issues, such as synergistic survey strategies, calibration networks, and spectroscopic follow-up.

\begin{displayquote}
\textbf{m. What resources, capabilities and infrastructure at DOE National Laboratories or the NASA Centers would be beneficial for and could accelerate or facilitate research in this topic?}
\end{displayquote}

DOE National Laboratories and the NASA Centers can jointly support mutually beneficial resources such as long-term computing infrastructure personnel, who have different skill sets and career paths than typical academics. This can be important to ensure the access to technically focused long-term joint processing and simulation efforts. Storage and computational time in the Rubin and \textit{Roman} analysis pipelines and user spaces to do these tasks. Many of the previously mentioned task forces can be coordinated out of one or more DOE National Lab or NASA Center.

\begin{displayquote}
\textbf{n. Are there other factors, not addressed by the questions above, which should be considered in planning HEP and APD activities in this subject area?}
\end{displayquote}

No answer.

\section{Conclusion}

Type Ia supernovae (\sne) are a key cosmological probe, and the supernovae that the three great observatories - \textit{Euclid}, \textit{Nancy Grace Roman Space Telescope} and the Vera C. Rubin Observatory - will discover promise to revolutionize our understanding of the nature of dark energy.
These transients are already key science drivers for two of the three large surveys of the 2020s: LSST and \textit{Roman}, which have large and active communities involved in all aspects of these missions. Combining data sets allows increases in redshift range and wavelength range, boosting the statistical precision and the systematic control.
We discussed improvements to both statistical and systematic precision that combinations of data sets will enable such as improved photometric calibration, \sn standardization, and redshift measurements. Other science cases will also benefit from these synergies. In particular, the photometric redshift improvements will benefit cosmological measurements from weak lensing. Also the discovery and confirmation of strong-lensing systems, galaxy clusters, and gravitational wave counterparts.

While the community will benefit from combining data from these observatories, orchestrating them to act in concert with one another will yield the greatest scientific benefits. Particularly due to the time-dependent nature of SN surveys, planning must be done both prior to and during the surveys. We have discussed what teams and plans should be put in place now to start preparing for these combined sets by setting up cross-agency task forces. These task forces will allow for the needed communication to improve survey operations, calibration, spectroscopic follow-up, joint pixel-processing, and analysis. By working together, the DOE and NASA, can improve their investments with an increase of scientific output, and in particular, a deeper understanding of the nature of dark energy and our Universe.

\vspace{3em}
\noindent The \textit{Roman} Supernova Science Investigation Teams and the LSST DESC Supernova Working Group endorse this response.

\bibliographystyle{apj}
\bibliography{library}


\end{document}